\documentclass[12pt]{iopart}
\usepackage[colorlinks]{hyperref}
\usepackage{amssymb,amsbsy,graphicx,color}

\makeatletter
\renewcommand\tableofcontents{\section*{\contentsname}\@starttoc{toc}}
\makeatother

\makeatletter
\def\@appendixstar{\@@par
 \ifnumbysec                         
 \@addtoreset{table}{section}        
 \@addtoreset{figure}{section}\fi    
 \setcounter{section}{0}
 \setcounter{subsection}{0}
 \setcounter{subsubsection}{0}
 \setcounter{equation}{0}
 \setcounter{figure}{0}
 \setcounter{table}{0}
 \def\thesection{\Alph{section}}
 \def\theequation{\ifnumbysec
      \Alph{section}.\arabic{equation}\else
      \Alph{section}\arabic{equation}\fi}  
 \def\thetable{\ifnumbysec                 
      \Alph{section}\arabic{table}\else
      A\arabic{table}\fi}
 \def\thefigure{\ifnumbysec
      \Alph{section}\arabic{figure}\else
      A\arabic{figure}\fi}}

\makeatother

\begin{document}

\newtheorem{theo}{Theorem}
\newtheorem{lemma}{Lemma} 

\title[Engineering States, Measurements and Anomalous Diffusion]{Engineering Quantum States, Nonlinear Measurements, and Anomalous Diffusion by Imaging}

\author{Kurt Jacobs}

\address{Department of Physics, University of Massachusetts at Boston, 100 Morrissey Blvd, Boston, MA 02125, USA}

\author{Daniel A. Steck} 

\address{Department of Physics and Oregon Center for Optics, 1274 University of Oregon, Eugene, OR 97403-1274}

\begin{abstract}
We show that well-separated quantum superposition states, measurements of strongly nonlinear observables, and quantum dynamics driven by anomalous diffusion can all be achieved for single atoms or molecules by imaging spontaneous photons that they emit via resonance florescence. To generate anomalous diffusion we introduce continuous measurements driven by L\'evy processes, and prove a number of results regarding their properties. In particular we present strong evidence that the only stable L\'evy density that can realize a strictly continuous measurement is the Gaussian. 
\end{abstract}

\pacs{42.50.Dv, 03.65.Ta, 02.50.Ey, 05.40.Fb}
\maketitle

\tableofcontents

\section{Introduction}

The measurement of the position of a particle is perhaps the most basic quantum-mechanical measurement, but it poses many subtleties. The archetype for a position measurement is the Heisenberg microscope---the imaged detection of light scattered from the particle. Here we consider photodetection of the light emitted by a single two-level atom. If the photons first pass through a lens designed to form an image of the atom before detection, then the photodetections provide information about the location of the atom. The detections therefore modify the spatial wave function of the atom.  
This situation was first considered by Holland \textit{et al.}~\cite{Holland96}, as a tool for the efficient simulation of atomic decoherence due to spontaneous emission. 
As applications, they considered spatial flights correlated with quantum jumps,
and L\'evy statistics due to trapping in potential wells of an optical lattice~\cite{Marksteiner96}.

Under the proper conditions, this type of imaged photodetection can lead to the standard form of a continuous position measurement~\cite{JacobsSteck06}, which gives rise to Gaussian projection noise~\cite{CMnotes}. However, as we will discuss in detail, it is the diffraction pattern of the imaging system that gives the form of the collapse in the Heisenberg microscope.  Imaging systems often involve hard-edged apertures and thus long-tailed (non-Gaussian) diffraction patterns. Such long-tailed distributions lead in the context of random walks to exotic L\'evy noise---so-called ``anomalous diffusion''~\cite{Klafter96, Tsallis95, Metzler00}. Another situation in which one can naturally tailor the form of the collapse would be the detection of resonance fluorescence in the presence of a magnetic field gradient~\cite{Thomas95}. In this case it is the Lorentzian line shape that enters as the collapse function. Thus, in moving towards experiments to realize continuous position measurements via imaging, it is important to understand the impact of long-tailed collapse operators, and under what conditions there may be qualitative modifications to the quantum noise and the conditioned dynamics of the continuously observed atom.

Our purpose here is threefold. In Sec.~\ref{sec::2} we examine the effect of imaging on the wave-function of a single atom, and show that this can be used to prepare non-classical states, and realize highly nonlinear measurements. In Sec.~\ref{sec::3}, in preparation for exploring how imaging can be used to generate anomalous diffusion, we examine continuous measurements that contain L\'evy noise. We are able to show that there are no truly continuous measurements that are driven only by stable L\'evy processes~\cite{JacobsSP, ContTankov}, with the sole exception being the Gaussian. It turns out that the underlying reason for this is a result that we prove in Sec.~\ref{sec::4}: repeated measurements whose errors are given by the stable L\'evy distributions will eventually collapse the wave-function to a Gaussian, just like the usual Gaussian measurements. 
In Sec.~\ref{sec::3} we also discuss how truly continuous measurements \textit{can} contain L\'evy noise, by combining a stable L\'evy process with a Poisson or Gaussian process. Finally, in Sec.~\ref{sec::4} we use the above results to examine how imaging can be designed to make measurements that induce anomalous diffusion, an effect that, in this case, is quantum-mechanical in origin.

\section{Position measurements via photodetection}
\label{sec::2}
\begin{figure}[t]
  \begin{center}
     \includegraphics[width=4in]{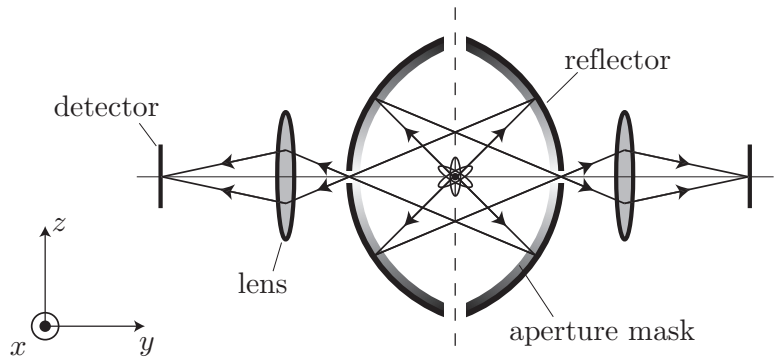}
  \end{center}
  \vspace{-5mm}
  \caption
         {Spontaneous emissions from a single atom or molecule are imaged by mirrors. Each mirror has a mask whose transmission profile can be modified. 
	\label{fig::setup}}
\end{figure}

We consider the motion of a single two-level atom (or molecule) along a single direction, which we will refer to as the $z$-axis. We illuminate the atom with a resonant laser traveling along the $x$-axis, and place two lenses or mirrors on opposite sides of the atom that image the emitted photons in the $xz$-plane. This configuration is depicted in Fig.~\ref{fig::setup}.  In addition to forming an image of the atom at the detector, the mirrors have a mask over them, which can apply an angle-dependent phase, as well as having an angle-dependent absorption profile. This mask introduces an aperture for the imaging optics. 

To describe the angular dependence of the aperture, we use spherical polar coordinates $\theta$ and $\phi$. As usual $\theta$ is the angle to the $z$-axis, and $\phi$ is the angle in the $xy$-plane. The center of each lens is on the $y$-axis, and thus given by $\theta = \pi/2$. We parameterize the distance from the center of each lens by $\xi = \cos\theta \in [-1,1]$.   We will denote the aperture transmission function of the aperture, $t(\xi,\phi)$, and for simplicity we will take this to be independent of $\phi$ in a fixed window $\Delta\phi$ (centered at the center of the lens), and to be zero otherwise.  We note that since the aperture may include a phase mask, $t$ can be complex.  
  
The basic analysis of the above imaging setup has been performed in Refs.~\cite{Holland96, JacobsSteck06}, and the relevant results are as follows. Upon detecting a single photon at location $z=a$ on the image, the wavefunction of the atom, $\psi(z)$, undergoes the transformation 
\begin{equation}
    \psi(z) \rightarrow \frac{A(z - a) \psi(z)}{\mathcal{N}} , 
\end{equation}
where $\mathcal{N}$ is the required normalization, and the ``collapse" operator $A(z)$ is 
\begin{equation}
       A(z)  =  \int_{-1}^{1} \!\!\! \chi (\xi) \, e^{ i k z \xi}  \, d\xi . 
       \label{eq::Az}
\end{equation}
Here $k$ is the wave number of the emitted photon, and 
\begin{equation}
       \chi (\xi) = t(\xi) \sqrt{f(\xi)}, 
\end{equation}
where $f(\xi) = (3/4)\sin^2(\theta)$ is the dipole angular emission function for the photon. By choosing the aperture transmission appropriately, $\chi$ can be {\em any} function satisfying $|\chi (\xi)|^2 \leq f(\xi)$, with the restriction that $\chi $ is zero outside the aperture capture region. The fraction $\eta$ of photons captured by the optics is given by the integral of $|\chi |^2$ over the capture region.  An emitted photon that is not detected transforms the atomic wave function via the collapse operator $B(z-a)$, where $B(z)$ is given by Eq.~(\ref{eq::Az}), with $\chi$ replaced with $\chi' = \sqrt{f}(1 - t)$. Since the photon is not detected, the density matrix of the system is given by averaging over all the possible image locations $a$. 

The power of the above imaging setup comes partly from the ability to engineer the collapse (or {\em measurement}) operator $A(z)$ by selecting an appropriate aperture profile, $t(\xi)$. Since $t(\xi)$ is zero outside $[-1,1]$, for a given $A(z)$ we can find the required profile by using the inverse Fourier transform: 
\begin{equation}
       t (\xi)  \propto  \frac{1}{\sqrt{f(\xi)}}  \int_{-\infty}^{\infty}  \!\!\!  A(z) \, e^{ -i k z \xi} \, dz . 
\end{equation} 
The other key component of imaging is the ability to engineer the mixing of optical modes prior to detection. We now present three examples that illustrate how these two components can be exploited. 

\subsection{Application: Preparing spatially separated superpositions} 
\begin{figure}[t]
  \begin{center}
     \includegraphics[scale=1]{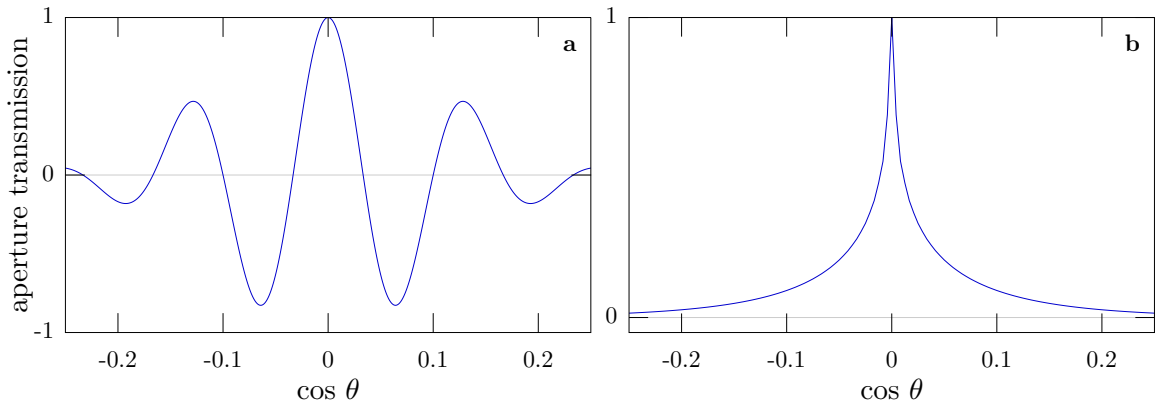}
  \end{center}
  \vspace{-5mm}
  \caption
         {Aperture amplitude transmission profiles for preparing (a) a spatially separated superposition, and (b) a state with an infinite position variance.  
	\label{fig::Tprof}}
\end{figure}

The imaging configuration described above can be used to prepare an atom in a superposition of two spatially separated states. This is achieved by chosing the aperture of the imaging system so that the collapse function $A(z)$ is the sum of two well-separated Gaussians. Let $\sigma^2$ be the variance of both Gaussians, and $L\gg\sigma$ their separation. An aperture function that does this is shown in Fig.~\ref{fig::Tprof}. This aperture covers about $29^\circ$ of arc in $\theta$ (i.e. $\xi \in [-1/4,1/4]$), and gives well-separated Gaussians with $\sigma = 1.5\lambda$ and $L = 15\lambda$. Here $\lambda$ is the wavelength of the emitted photon. The separation is easily increased by increasing the number of oscillations in the profile. The fraction of photons captured by the aperture, assuming the two-mirror configuration in Fig.~\ref{fig::setup}, and $\Delta\phi = 29^\circ$, is approximately 1/186. The aperture size we have chosen is conservative --- using two mirrors with $120^{\mbox{\tiny o}}$ of arc as the focussing elements, capture rates greater than 1 in 20 should be feasible with this aperture profile. 

One now prepares the atom in a broad Gaussian state, centered at $z=0$, by cooling it to the ground state of a harmonic trap using standard techniques~\cite{Diedrich89}. One then illuminates the atom with a laser pulse to prepare it in the excited state, and allows it to emit a photon. This preparation process is repeated until the emitted photon is detected. (One must repeat the preparation because a single undetected emission destroys the spatial coherence.) This prepares the atomic wave function in a superposition of two spatially separated wave-packets, with separation $L$ and widths $\simeq \sigma$. The relative heights of the two localized wave-packets is determined by the location of the detected photon. The closer the photon to $z=0$, the more equal the weighting of the two packets. Naturally one requires that the probability of the smaller packet, $P_{\mbox{\scriptsize s}}$, should not be too small. If the ground state of the trap is broad compared to $L$, a numerical calculation shows that when a photon is detected, the probability that it gives $P_{\mbox{\scriptsize s}} \geq 1/3$ is approximately $95\%$. 
Furthermore, the superposition can be made arbitrarily symmetric with a sufficient number of repetitions of the preparation process.

\subsection{Application: Measuring nonlinear observables} 

Measurements of nonlinear observables have applications in state-preparation~\cite{Jacobs09} and in detecting signatures of quantum motion, among others. The usual method of measuring nonlinear observables (e.g. nonlinear functions of position, $z$), is to couple a probe to the system via a nonlinear interaction proportional to the observable. Generating effective nonlinear interactions is not easy, however. Feasible methods exist to generate effective $z^4$ interactions, but generating higher powers becomes increasingly more difficult~\cite{Jacobs09c}. We show now that imaging can be used to engineer a measurement of the highly nonlinear function, $|z|$, without the need for a nonlinear interaction. This measurement can be used to generate spatially separated superposition states, in a similar manner to that of $z^2$ measurements~\cite{Jacobs09}, with the advantage that it is linear everywhere but at $z=0$. This linearity means that the wave-packets in the resulting superposition will be very nearly Gaussian squeezed states.  

To realize a measurement of $|z|$ we make the aperture fully transmitting. This allows one, in principle, to capture the majority of emitted photons. This time we use the optics to first collimate the light. We then split the light into two beams with a 50-50 beam splitter, invert one of them about $z=0$, and recombine them. Finally we focus the beam to an image, which is now a superposition of the original image and its reflection about $z=0$. Since we cannot now tell $z$ from $-z$, the detected photon only contains information about $|z|$.  This measurement can be made quasi-continuous by repeatedly exciting the atom and detecting the emitted photons. The method can also be extended to more complex measurements, erasing information about multiple intervals of the values of $z$. 

\section{Anomalous diffusion and L\'evy measurements} 
\label{sec::3}
All continuous measurements that have been realized to date on quantum systems generate Gaussian or Poissonian noise. That is, in each infinitesimal time step $dt$, the error in the measurement record, and the noise induced in the system, is either the Gaussian (Wiener) increment $dW$, with variance $dt$, or the discrete jump of a Poisson process. These noise processes are ubiquitous in nature because of the central limit theorem. There are, however, a whole range of more exotic noise processes that do not have these statistics, and find applications in wide range of phenomena~\cite{Klafter96, Tsallis95}. A remarkable class of these are the {\em symmetric stable L\'evy processes}, $L_\alpha(t)$, indexed by the continuous parameter $\alpha \in (0,2]$~\cite{JacobsSP, ContTankov}. We will denote the random infinitesimal increment of one of these processes by $dL_{\alpha}$. When $\alpha =2$, the L\'evy process is just the Wiener process. All other stable L\'evy processes break the central limit theorem because they have infinite variances. 

The ``stable'' L\'evy processes are so-called because the sum of two or more of their increments has the {\em same} probability density as each individual increment (except that the width of this density is larger by a factor of $2^{1/\alpha}$). This is familiar for Gaussian random variables (the sum of two Gaussian deviates is also Gaussian), but is usually not true for other probability densities. In fact, if a probability density satisfies the conditions for the central limit theorem, and is not Gaussian, then the sum of multiple samples from this density \textit{must} differ from the original density, since ultimately this sum must have a Gaussian density in the limit of many samples. 

\begin{figure}[t]
  \begin{center}
     \includegraphics[width=4in]{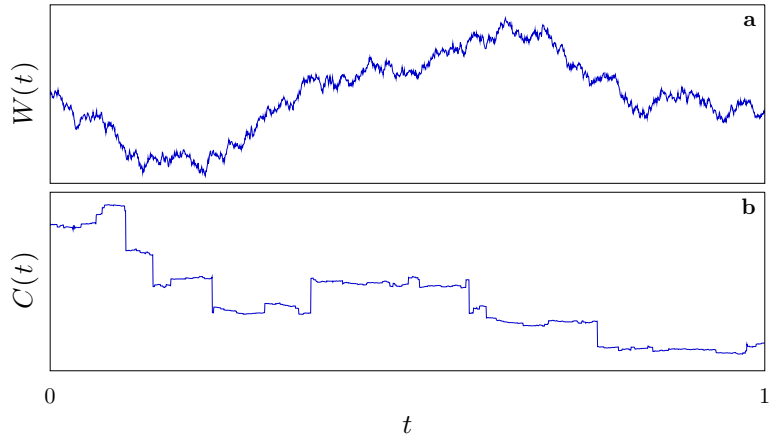}
  \end{center}
  \vspace{-5mm}
  \caption
         {(a) Wiener noise and (b) Cauchy noise.   
	\label{fig:paths}}
\end{figure}

When two increments of the (Gaussian) Wiener process are added together, the width (uncertainty) of the sum is $\sqrt{2}$ times that of each increment. As stated above, for the stable L\'{e}vy process $L_\alpha$ this factor becomes $2^{1/\alpha}$. This factor determines how the width (the uncertainty) of the process scales with time. The result is that the process $L_\alpha (t)$ has a width proportional to $t^{1/\alpha}$. For Gaussian noise this reduces to the familiar scaling: the standard deviation increases as $\sqrt{t}$. This is the usual behavior of diffusion, for example the diffusion of a pollen grain undergoing Brownian motion. If the uncertainty of a dynamical system scales instead as $t^{1/\alpha}$, for $\alpha < 2$, then this behavior is referred to as {\em anomalous diffusion}~\cite{Metzler00}. The case $\alpha=1$ is the Cauchy process, whose increment, $dC \equiv dL_1$, has the {\em Lorentzian} probability density 
\begin{equation}
   P(dC) = \frac{dt}{\pi[(dC)^2 + (dt)^2]} . 
\end{equation}
The uncertainty of the Cauchy process scales at $t$. For readers new to L\'{e}vy processes, realizations of the Wiener and Cauchy processes are displayed in Fig.~\ref{fig:paths}. A realization, or ``sample path'', of the Cauchy process is quite distinct from the Wiener process. An introduction to L\'{e}vy processes may be found in~\cite{JacobsSP, ContTankov}. With this background out of the way, we ask how stable L\'{e}vy processes might be realized by continuous quantum measurements. 

\subsection{Absence of continuous measurements with stable L\'evy noise} 
           
A continuous measurement of position, $X$, is characterized by the error of the measurement result in an infinitesimal time-interval $dt$. Specifically, if $d\varepsilon$ is the random variable describing this error, then the increment of the measurement result obtained in the time-interval $dt$ is given by 
\begin{equation}
   dr =  \langle X \rangle dt + \gamma\, d\varepsilon, 
   \label{eq::dr}
\end{equation}
where $\gamma$ is a constant that fixes the overall size of the error. Gaussian measurements have $d\varepsilon = dW$~\cite{JacobsSteck06, WMBook, Brun02}. Little is yet known as to what classes of exotic L\'{e}vy processes can appear as the errors in continuous measurements~\cite{Holevo86}. We now show that continuous measurements can \textit{never} have errors given purely by stable L\'evy processes, with the sole exception being the Gaussian. 

The amount of information that the measurement obtains about $X$ in a short time interval $\Delta t$ is determined by the width of the probability density of the measurement result in each time interval.  The measurement result is given by dividing $dr$ by $dt$, and is the mean value of the measured quantity plus the random error: 
\begin{equation}
   \frac{dr}{dt} =  \langle X \rangle + \gamma\, \frac{d\varepsilon}{dt}. 
\end{equation}
The amount of information extracted is determined by the width of the probability density of $d\varepsilon/dt$; if the error has a large ``variance'' (width) the observer gains little information, and vice versa. In particular, the observers probability density for $X$ (the diagonal elements of the density matrix in the basis of $X$) after the measurement is given by multiplying her initial probability density by the probability density for $d\varepsilon/dt$, shifted by the measurement result, $dr/dt$~\cite{JacobsSteck06}. 

For a sequence of measurements to produce a valid continuous measurement in the continuum limit, the amount of information extracted (the change to the observers state-of-knowledge) must scale in the right way as we reduce the time interval between the measurements, $\Delta t$. In particular, the average change to the observer's state-of-knowledge in a fixed time interval, $T$, must remain the same as we reduce the time-step and thus increase the number of measurements in the interval. For this reason we can show immediately that Eq.~(\ref{eq::dr}) cannot describe a continuous measurement if $d\varepsilon = dL_\alpha$ for $\alpha \leq 1$. Consider first the case when $\alpha = 1$. Because the width of $dL_1$ scales as $dt$, we see that the width of $d\varepsilon/dt = dL_1/dt$ is \textit{independent of dt}. Since it does not decrease with $dt$, as we increase the number of measurements, and take the limit $dt \rightarrow 0$, the amount of information extracted by the observer tends to infinity, resulting in an instantaneous collapse of the wave-function. This will also be true for $\alpha < 1$. 

\begin{figure}[t]
  \begin{center}
     \includegraphics[scale=1]{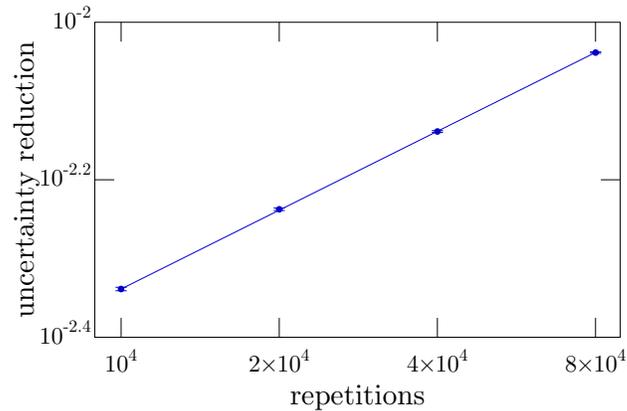}
  \end{center}
  \vspace{-5mm}
  \caption
         {The scaling of the average reduction in an observer's uncertainty, for a sequence of stable L\'evy measurements, as the time step is decreased, and thus as the number of repetitions is increased. The horizontal axis gives the number of repetitions. The straight line gives the best (logarithmic) fit. 
	\label{fig:stable_test}}
\end{figure}

Showing that continuous measurements cannot be realized for $1 < \alpha < 2$ is not so simple, since in this case the width of the probability density for the error, $d\varepsilon/dt = dL_\alpha/dt$, 
diverges as $dt \longrightarrow 0$, and thus the extracted information tends to zero as required. To determine whether a continuum limit exists for these measurements we 
begin by way of a numerical example before we proceed with an analytic calculation.
We start with a state of knowledge for $X$ given by a Gaussian with unit variance, and simulate a sequence of measurements for $\alpha = 1.5$. This involves repeatedly multiplying by the probability density for $L_\alpha$, which must be obtained from the Fourier transform of its characteristic function~\cite{JacobsSP}. We simulate four sequences of measurements, in which each sequence has half the time-step of the one before. We then examine how the change in the width of the observer's state-of-knowledge scales with the time-step. We obtain this change by averaging each sequence over four thousand realizations. The results for the four sequences are displayed on a log-log plot in Fig.~\ref{fig:stable_test}.  From this we see that the reduction in the observers uncertainty increases as a power of the inverse time-step (the exponent, or slope in the plot here, is consistent with $1/3$),
and thus will reduce the observer's uncertainty to zero in the limit $dt \rightarrow 0$. 
A sensible continuum limit would require a flat slope (flat scaling exponent) in such a plot as this,
as happens in the Gaussian case $\alpha=2$.  Again, this means that the amount of information
extracted per unit time becomes asymptotically constant as $dt\longrightarrow 0$.

To test this for all $\alpha\in (1,2)$, we can compute the scaling exponent as a function of $\alpha$. To perform this calculation numerically, we can use a deterministic method that is more efficient than the Monte Carlo method of Fig.~\ref{fig:stable_test}. The idea is to take a narrow, Gaussian initial state, and compute the uncertainty reduction that results from applying the measurement operator centered about a particular measurement result. We then average the reduction in uncertainty over all measurement results by performing a numerical integral. 

The above numerical procedure for calculating the scaling exponent as a function of $\alpha$ is rather cumbersome, since there is not a general expression for the measurement operator for arbitrary $\alpha$. Remarkably we can bypass this procedure, and obtain a simple analytic expression for this scaling exponent, using the following insight: if the initial state is Gaussian, the application of the measurement function (L\'evy density) \textit{leaves the final state in a Gaussian distribution, and the reduction of the variance behaves, on average, as if the measurement function were itself Gaussian}.  This is not an obvious statement, and we will justify it in Sections \ref{section:arbproject}, \ref{section:gaussian}, and Appendix~\ref{section:appendix-clt}. This result means that the reduction in the variance due to the collapse is simple for every L\'evy measurement: it is proportional to the inverse square of the width of the distribution for the error). Now that we know how the reduction in the variance depends on the width, and we know how the width of the L\'evy distribution scales with $dt$, we can easily determine the scaling of the reduction in the variance. Specifically, if the measurement error is given by $dL_\alpha/dt$, then the square width of this distribution is $w^2(dt) = dt^{2/\alpha-2}$. So the reduction in the variance for a single measurement for duration $dt$ is $\Delta V(dt) \propto - 1/(w^2) = - dt^{2 - 2/\alpha}$. If we now make $N$ measurements, each with duration $dt/N$, then the total reduction is 
\begin{equation} 
   \Delta V_{\mbox{\scriptsize tot}} =  N \Delta V (dt/N)  \propto  - N (dt/N)^{2 - 2/\alpha} = N^{2/\alpha-1} \Delta V(dt). 
\end{equation} 
The scaling exponent we seek is therefore $2/\alpha-1$. 

\begin{figure}[t]
  \begin{center}
     \includegraphics[scale=1]{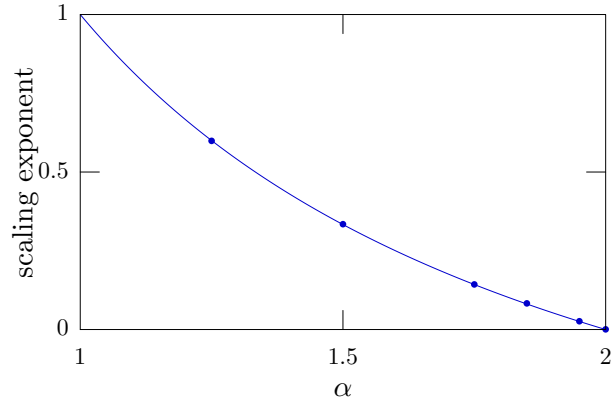}
  \end{center}
  \vspace{-5mm}
  \caption
         {The dependence of the power-law scaling exponent
(i.e., the slope of the line in Fig.~\ref{fig:stable_test}) as a function of 
$\alpha$.  The solid curve is the analytic result ($2/\alpha-1$),
  and the points indicate values obtained by a numerical calculation of the 
  uncertainty reduction due to projection with the corresponding L\'evy distribution.
	\label{fig:alphascaling}}
\end{figure} 

Fig.~\ref{fig:alphascaling} shows the dependence of the power-law scaling exponent
(i.e., the slope of the line in Fig.~\ref{fig:stable_test}) as a function of $\alpha$.  The analytic result $2/\alpha-1$ is shown along with numerical tests at the specific points indicated. Again, only a scaling exponent of zero can correspond to a sensible continuous limit of a sequence of L\'evy measurements.  
Thus we see that, even in the range $1<\alpha<2$, the uncertainty decreases \textit{too quickly} to correspond to a continuous measurement, as the continuous limit will lead to instantaneous collapse of the state---essentially, corresponding to a divergent measurement strength. Though we have only analyzed a Gaussian state, a valid continuous measurement must work for \textit{every} state, and thus our analysis is sufficient to rule out these kinds of continuous measurements.  (Further, as we will show below, measurements with L\'evy collapse operators tend to produce Gaussian states anyway.)
Consequently, only the Gaussian ($\alpha=2$) case can correspond to a continuous measurement.

\subsection{Chained processes and exotic noise}

Despite the above result, it {\em is} possible to construct continuous measurements that contain the exotic statistics of {\em any} stable L\'evy process. To do so we combine the stable process with a Poisson process in the following way:  we replace the time index upon which the stable process $L_\alpha(t)$ depends with the Poisson process ${\mathcal P}(t)$, to form the ``chained'' process ${\mathcal S}_\alpha(t) = L_\alpha({\mathcal P}(t))$. (In the mathematics literature this chaining procedure is referred to as {\em subordinating} $L_\alpha$ with ${\mathcal P}$ \cite{ContTankov}.) Now examine the behavior of this new L\'evy process. Between jumps of the Poission process the time index for the Cauchy process does not change, and thus the increment $d{\mathcal S}_\alpha$ is zero. Upon an event (a jump) in the Poisson process, the time index of the stable process increases by unity, and thus generates a finite increment $\Delta L_\alpha = \int_0^1 dL_\alpha$. The temporal scaling of the stable process $L_\alpha$ no longer prevents the process ${\mathcal S}_\alpha$ from representing a continuous measurement, since the scaling as $\Delta t \rightarrow 0$ is entirely determined by the Poisson process. 

When there are many jumps in a given time interval, the number of jumps fluctuates only a little from its mean value. Because of this, if we make the rate of jumps $\lambda$ large, and set $\sigma \propto 1/\lambda$ so that the effect of each jump on the measured system is small, then the Poisson-subordinated L\'evy measurement realizes what is effectively a quasi-continuous L\'evy measurement, with $\Delta t = 1/\lambda$. Our simulations show that the measurement records of these measurements are qualitatively the same as those of the stable L\'evy processes (e.g., Fig~\ref{fig:paths}). 


Measurements described by the L\'evy processes $\mathcal{S}_\alpha$ can be realized by using the imaging setup of Fig.~\ref{fig::setup}. The photodetection events are the jumps of the Poisson process,  so we need each of these to correspond to a measurement with the result $\Delta r = \langle X \rangle + \gamma \Delta L_\alpha$ for some $\gamma$. This is accomplished by choosing the aperture so that the square of the collapse operator is the density $P(\Delta L_\alpha)$, with $\gamma$ set by scaling the width of $P$. In Fig.~\ref{fig::Tprof}b we give the aperture profile $T(\xi)$ for a Cauchy measurement. This prepares a state with a Cauchy probability density for position, and we display the Wigner function for this state in Fig.~\ref{fig:Wig}. 

\section{Anomalous diffusion and continuous measurement}
\label{sec::4}

\subsection{Action of an arbitrary position measurement}\label{section:arbproject}
So far we have focused on the statistics of the measurement results. Now we turn to the question of generating anomalous diffusion in the {\em dynamics} of a quantum system. 
In particular, we address the question of what happens to the quantum state under the action
of a L\'evy (or any other) spatial projection operator.
We will assume a Gaussian initial state, with probability density
\begin{equation}
  \rho(x,x) = \frac{1}{\sqrt{2\pi}\sigma}e^{-x^2/4\sigma^2}, 
\end{equation}
where for convenience we assume the state is centered at $x=0$.
As we are interested in the 
quasi-continuous---and thus weak-measurement---limit, we will assume that the 
width of the projection operator is much
wider than the width of the state.
Given a Hermitian collapse operator $\Omega$, we have the 
(unnormalized) measurement reduction 
$\rho\longrightarrow\Omega\rho\Omega$,
or
$\rho(x,x)\longrightarrow\Omega^2(x)\rho(x,x)$ in the position representation.
The operator $\Omega^2(x)$ is very broad, though not necessarily centered at $x=0$.
Given that it varies slowly over the extent of the state, we can expand the operator
to second order as
\begin{equation}
  \Omega^2(x) 
  = \Omega_0+\Omega_1 x+\Omega_2 x^2+O(x^3),
  = \Omega_0e^{ax-bx^2}+O(x^3)
\end{equation}
provided that $a=\Omega_1/\Omega_0$ and $b=a^2/2 + \Omega_2/\Omega_0$.
After dropping normalization factors, the reduced state is
\begin{equation}
  \Omega^2(x) \rho(x)\propto
  e^{ax-bx^2}e^{-x^2/4\sigma^2}\propto e^{-(x-\mu)^2/4\tau^2},
\end{equation}
where 
\begin{equation}
  \mu=\frac{2a\sigma^2}{1+4b\sigma^2},\qquad \tau=\frac{\sigma}{\sqrt{1+4b\sigma^2}}
\end{equation}
are the new mean and standard deviation, respectively,

There are several points to note here.
For a Gaussian collapse operator, for example, the $b$ parameter is \textit{constant},
since it is only the $a$ parameter that controls the displacement of the Gaussian from
$x=0$. This reiterates the well-known result that a Gaussian measurement will
reduce the uncertainty of a Gaussian state \textit{deterministically}---i.e., independent of the
measurement result.
In particular, $b=1/2\sigma'^2$, where $\sigma'$ is the standard deviation of the 
measurement function.
For any \textit{other} form of the collapse operator, a particular measurement may
increase or decrease the uncertainty.  It is only on \textit{average} that the 
uncertainty of the state decreases.
Again, the $b$ parameter, which represents a particular function of the 
local curvature $\Omega_2$ of the collapse function at $\langle x \rangle$,
controls how the uncertainty changes due to the collapse.
For example, for a long-tailed distribution such as the Cauchy, the uncertainty 
\textit{decreases} if the measurement result is close to zero, but \textit{increases}
if the measurement result is very far from zero.
Also since the $a$ parameter controls the displacement of the measurement operator,
it likewise controls the shift of the mean of the state in response to the measurement,
as we see from $\mu\propto a$.
However, the main point of this section is this: in this regime of broad measurement
distributions, the action of the collapse preserves the Gaussian form of the quantum state,
\textit{independent} of the form of the measurement operator.
In particular, even measurements of stable L\'evy distributions preserve the Gaussian
form of the quantum state.

\subsection{Collapse to a Gaussian wave packet}\label{section:gaussian}

Now we extend the analysis above and prove a somewhat surprising result:
\begin{figure}[t]
  \begin{center}
     \includegraphics[scale=1]{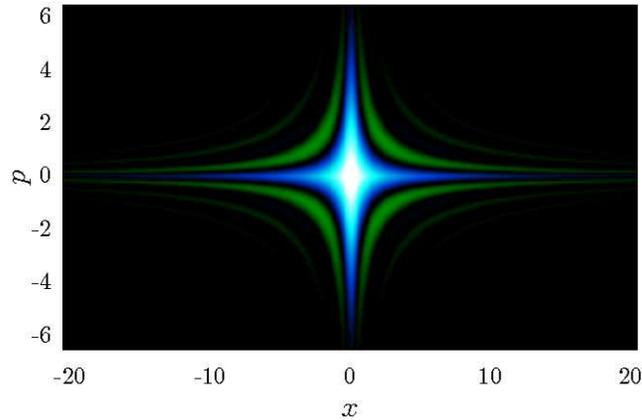}
  \end{center}
  \vspace{-5mm}
  \caption
         {Wigner function for a state with a Cauchy probability density for position. Here $X$ and $P$ are scaled so that $[X,P] = i$, and the width of the Cauchy density is $\sigma = 0.3$. Blue is positive, green negative.   
         \label{fig:Wig}}
\end{figure} 

{\em Theorem:} In the absence of any Hamiltonian evolution, a sequence of many repeated L\'evy measurements generates a Gaussian wave function --- independent of the 
initial state --- just as do Gaussian measurements. 

{\em Proof:} A sequence of $N$ L\'evy measurements of $X$ corresponds to multiplying the initial position density, $P(X) = |\psi(x)|^2$, by a sequence of L\'evy densities. This means that the characteristic function of the resulting density is the convolution of the characteristic functions of $N$ L\'evy densities. The characteristic functions of the symmetric $\alpha$-stable densities, $L_\alpha(x)$, are $\chi_\alpha(s) = \exp[-\sigma^\alpha |s|^\alpha  + i\mu s]$, where $\sigma$ is a width parameter and $\mu$ is the mean \cite{JacobsSP, ContTankov}. For $\alpha\geq 1$, $\mu$ is the mean of the L\'evy density. If we now think of $\chi_\alpha(s)$ as a probability density (for $s$), then convolving $N$ of them corresponds to adding $N$ random variables with density $\chi_\alpha$. Since every $\chi_\alpha(s)$ has a finite second moment, the central limit theorem tells us that for large $N$ the result of these convolutions tends to a Gaussian. (Note that the usual central limit theorem does not handle complex-valued distributions, so we extend the theorem to cover this case in Appendix~\ref{section:appendix-clt}.) If the final characteristic function is Gaussian, then the final density is also Gaussian. Thus, while a {\em single} L\'evy measurement on an initially flat density creates a L\'evy density, many repeated L\'evy measurements generate Gaussian densities, and thus Gaussian wave functions. $\square$

\subsection{Anomalous diffusion induced by back-action: a proposed experiment}

Once the wave-function of a single particle has been reduced to a Gaussian, even the exotic measurements discussed above induce only Gaussian noise in the dynamics of $X$. This is not difficult to show analytically for very strong (small $\gamma$) and very weak (large $\gamma$) measurements, and the numerical simulation we have performed confirm it more generally.  So how does one make measurements that {\em do} generate anomalous diffusion?  We now show that one can do this using imaging, and in a manner that directly exploits the quantum nature of the measurement back-action. This distinguishes it from the L\'evy noise in Ref.~\cite{Marksteiner96} (as well as the momentum L\'evy flights in Refs.~\cite{Bardou94, Bardou01, Breuer07} that arise in subrecoil laser cooling), which can be treated semiclassically. After initially preparing the atom in a broad Gaussian state, we make a ``square" measurement producing a wave function with  a square profile; that is, a flat wave function with a sharp cut-off at each end. This is achieved by choosing the aperture to be a sinc function. The quantum back-action of this measurement simultaneously generates a momentum probability density that is the square of the sinc function. This has an infinite variance. We then wait for a quarter period of the harmonic evolution, and this transfers the momentum density to position. Repeating the ``square-profile" measurement now produces an infinite-variance change in the mean position of the atom, and thus anomalous diffusion. Of course, in an experiment, each time we make a measurement, a number of photons will be lost before one is detected. We will refer to the average number of photons emitted for each one detected as the {\em loss rate}. We must check that the anomalous diffusion remains in the presence of this loss.  

\begin{figure}[t]
  \begin{center}
     \includegraphics[scale=1.0]{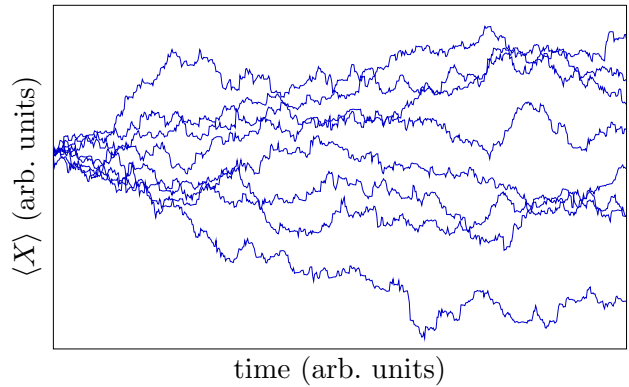}
  \end{center}
  \vspace{-5mm}
  \caption
         {Eight trajectories of the mean atomic position, resulting from photodetection using an aperture with a square profile, and total efficiency of 50\%.  
         \label{fig:eighttraj}}
\end{figure}

We now simulate this experiment, assuming mirrors with a $130^{\mbox{\tiny o}}$ arc. Using an aperture profile that gives a square-profile measurement with a width of $32$ wavelengths results in a loss rate of 119. We note that the loss rate per measurement can be reduced by post-selecting those photodetections in which the photo was detected very soon after the laser pulse. For these detections there will be, on average, many fewer photons that were not detected. For example, in the present case post-selecting $1$ in $118$ measurements reduces the loss rate per post-selected measurement to unity. 

To show that our square-profile measurement induces anomalous diffusion in the mean position, $\langle X \rangle$, we must examine the statistics of the change to $\langle X \rangle$ caused by a single photodetection. Let us denote the change induced in $\langle X \rangle$ as $\Delta \langle X\rangle$. To determine whether this change exhibits anomalous diffusion, we examine how the width of the probability density for $\Delta \langle X \rangle$, which we will call $\sigma_1$, compares to that for the sum of two independent samples of $\Delta \langle X \rangle$, which we will call $\sigma_2$. Anomalous diffusion is achieved when $\sigma_2 = 2^{\beta}\sigma_1$, with $\beta > 0.5$. To obtain accurate estimates of $\beta$ we must simulate sequences with  a large number of measurements. As a example, to determine $\beta$ when the loss rate per measurement is 5, we simulated 24 sequences of 600 photodetections each. This is numerically intensive due to the long tails of the wave functions involved, and so we employ a parallel computer. For a loss rate of unity (50\% efficiency) we find that $\beta = 0.69 \pm 0.16$, and for a loss rate of $5$ (17\% efficiency) $\beta = 0.60 \pm 0.07$. This shows us that while $\beta$ does decrease with increasing loss, the signature of anomalous diffusion remains. This indicates that one can generate exotic noise for considerably higher loss rates, although confirming this numerically is prohibitive with our present computing resources. Examples of the evolution of the mean atomic position, for a loss rate of 50\%, are shown in Fig.~\ref{fig:eighttraj}.  

\section{Acknowledgements} 
We thank Jeremy Thorn for comments and corrections. This work was performed with the supercomputing facilities in the College of Science and Mathematics at UMass Boston. K.J.\ is supported by the National Science Foundation under Project No.\ PHY-0902906, and D.A.S.\ is supported by the National Science Foundation under Project No.\ PHY-0547926. 

\appendix
\section{Appendix: Modified Central Limit Theorem}\label{section:appendix-clt}

The usual version of the central limit theorem states that, given a random
walk of independent, identically distributed steps, the resulting
distribution converges to a Gaussian in the limit of many steps $N$, with
a width scaling as $\sqrt{N}$.  The variance of each step must be finite for the
central limit theorem to hold.  Under certain other conditions, the one-step distributions
need not be identical. To support our argument in Section~\ref{section:gaussian},
we will now modify the standard proof (see, e.g., Ref.~\cite{Korner}) of the central limit theorem,
where the one-step distribution is the L\'evy characteristic function
\begin{equation}
  \chi_{\alpha,n}(s) = e^{-\sigma^\alpha |s|^\alpha  + i\mu_n s},
  \label{charfunconestep}
\end{equation}
where $\mu_n$ is randomly chosen on each ``step'' from the corresponding L\'evy
probability distribution $P_\alpha(x)$ with zero mean.
Obviously, this is not a sensible probability distribution. However, the central limit
theorem examines the successive convolutions of the one-step distribution, which is
well-defined, and ultimately what we are interested in.
The characteristic function (Fourier transform) of the one-step distribution 
(\ref{charfunconestep}) is
\begin{equation}
  \tilde{\chi}_{\alpha,n}(k) = 1-ik\tilde\mu_n
    -\frac{k^2(\tilde{\sigma}_n^{\,2}+\tilde{\mu}_n^{\,2})}{2}+O(k^3),
  \label{charfunconestepchar}
\end{equation}
where we have introduced an overall scaling factor to normalize the one-step distribution,
and the first two ``cumulants'' are given by
\begin{equation}
  \renewcommand{\arraycolsep}{0ex}
  \begin{array}{rcl}
    \tilde{\mu}_n &{}:={}&\displaystyle \int ds\,s\chi_{\alpha,n}(s)\\
    \tilde{\sigma}_n &{}:={}&\displaystyle \int ds\,(s-\tilde{\mu}_n)^2\chi_{\alpha,n}(s).
  \end{array}
\end{equation}
(Recall that the terms in the expansion of the characteristic function give the 
successive moments of the original distribution.) In general, both of these quantities
are nonzero, but the integrals are finite.
Also, $\tilde{\mu}_n$ is a random variable with zero mean and finite variance.

Now we will consider $N$ ``steps'' taken in the random walk, but first it is convenient to 
rescale the width of the one-step distribution to make it narrower by a factor
of $\sqrt{N}$.  This amounts to the replacements $s\longrightarrow s\sqrt{N}$
and $k\longrightarrow k/\sqrt{N}$, so that the characteristic function of the
characteristic function becomes
\begin{equation}
  \tilde{\chi}_{\alpha,n}'(k) = 1-\frac{ik\tilde\mu_n}{\sqrt{N}}
    -\frac{k^2(\tilde{\sigma}_n^{\,2}+\tilde{\mu}_n^{\,2})}{2N}+O(N^{-3/2}).
  \label{charfunconestepcharrescaled}
\end{equation}
After $N$ steps in the ``random walk,'' or the successive convolution of $N$ of
the one-step distributions, we have simply the product of the corresponding characteristic
functions via the convolution theorem (again dropping overall factors):
\begin{equation}
  \tilde{\chi}_{\alpha}^{(N)}(k) = \prod_{n=1}^N\left[1-\frac{ik\tilde\mu_n}{\sqrt{N}}
    -\frac{k^2(\tilde{\sigma}_n^{\,2}+\tilde{\mu}_n^{\,2})}{2N}+O(N^{-3/2})\right].
  \label{charfunconestepcharrescaledNsteps}
\end{equation}
The logarithm of the characteristic function of the convolution is
\begin{equation}
  \log\tilde{\chi}_{\alpha}^{(N)} = \sum_{n=1}^N\log\left[1-\frac{ik\tilde\mu_n}{\sqrt{N}}
    -\frac{k^2(\tilde{\sigma}_n^{\,2}+\tilde{\mu}_n^{\,2})}{2N}+O(N^{-3/2})\right].
  \label{charfunconestepcharrescaledNstepslog}
\end{equation}
Expanding the logarithm gives
\begin{equation}
  \renewcommand{\arraycolsep}{0ex}
  \begin{array}{rcl}
  \log\tilde{\chi}_{\alpha}^{(N)}
    &{}={}& \displaystyle \sum_{n=1}^N\left[-\frac{ik\tilde\mu_n}{\sqrt{N}}
    -\frac{k^2\tilde{\sigma}_n^{\,2}}{2N}+O(N^{-3/2})\right]\\
    &{}={}& \displaystyle -\frac{ik}{\sqrt{N}}\sum_{n=1}^N\tilde\mu_n
    -\frac{k^2}{2N}\sum_{n=1}^N\tilde{\sigma}_n^{\,2}.
  \end{array}
  \label{charfunconestepcharrescaledNstepslogexp}
\end{equation}
The usual central limit theorem applies to the first sum, which is $O(\sqrt{N})$,
so the first term remains finite with unit probability as $N\longrightarrow\infty$.
In the same way, the stochastic part of the second sum vanishes as $N\longrightarrow\infty$,
and the second term becomes $-k^2\langle\tilde{\sigma}_n^{\,2}\rangle/2$, where the angle
brackets denote an expectation value with respect to $\mu_n$.
Letting $k\longrightarrow -ik$, we obtain the cumulant-generating function for the
distribution after $N$ steps.  We see that in the limit of large $N$, 
the first two cumulants are finite and the rest vanish---this is the cumulant-generating
function for a Gaussian distribution.

\section*{References}


\begin{thebibliography}{10}

\bibitem{Holland96}
M.~Holland, S.~Marksteiner, P.~Marte, and P.~Zoller.
\newblock Measurement induced localization from spontaneous decay.
\newblock {\em {Phys.\ Rev.\ Lett.}}, 76:3683, 1996.

\bibitem{Marksteiner96}
S.~Marksteiner, K.~Ellinger, and P.~Zoller.
\newblock Anomalous diffusion and {L\'e}vy walks in optical lattices.
\newblock {\em Phys. Rev. A}, 53(5):3409--3430, 1996.

\bibitem{JacobsSteck06}
K.~Jacobs and D.~A. Steck.
\newblock A straightforward introduction to continuous quantum measurement.
\newblock {\em Contemp. Phys.}, 47:279, 2006.

\bibitem{CMnotes}
C.~M. Caves and G.~J. Milburn.
\newblock {\em Phys.\ Rev.\ A}, 36:5543, 1987.

\bibitem{Klafter96}
J.~Klafter, M.F. Shlesinger, and G.~Zumofen.
\newblock Beyond brownian motion.
\newblock {\em Phys. Today}, 49:33, 1996.

\bibitem{Tsallis95}
C.~Tsallis, S.~V.~F. Levy, A.~M.~C. Souza, and R.~Maynard.
\newblock Statistical-mechanical foundation of the ubiquity of {L}\'evy
  distributions in nature.
\newblock {\em Phys. Rev. Lett.}, 75:3589--3593, 1995.

\bibitem{Metzler00}
R.~Metzler and J.~Klafter.
\newblock The random walk's guide to anomalous diffusion: A fractional dynamics
  approach.
\newblock {\em Phys. Rep.}, 339:1--77, 2000.

\bibitem{Thomas95}
J.~E. Thomas and L.~J. Wang.
\newblock Precision position measurement of moving atoms.
\newblock {\em Phys. Rep.}, 262:311--366, 1995.

\bibitem{JacobsSP}
K.~Jacobs.
\newblock {\em Stochastic Processes for Physicists: Understanding Noisy
  Systems}.
\newblock CUP, Cambridge, 2010.

\bibitem{ContTankov}
R.~Cont and P.~Tankov.
\newblock {\em Financial Modelling with Jump Processes}.
\newblock Chapman \& Hall, New York, 2004.

\bibitem{Diedrich89}
F.~Diedrich, J.~C. Bergquist, W.~M. Itano, and D.~J. Wineland.
\newblock Laser cooling to the zero point energy of motion.
\newblock {\em {Phys.\ Rev.\ Lett.}}, 62:403, 1989.

\bibitem{Jacobs09}
K.~Jacobs, L.~Tian, and J.~Finn.
\newblock Engineering superposition states and tailored probes for
  nanoresonators via open-loop control.
\newblock {\em Phys. Rev. Lett.}, 102:057208, 2009.

\bibitem{Jacobs09c}
K.~Jacobs and A.~J. Landahl.
\newblock Engineering giant nonlinearities in quantum nanosystems.
\newblock {\em Phys. Rev. Lett.}, 103:067201, 2009.

\bibitem{WMBook}
H.~M. Wiseman and G.~J. Milburn.
\newblock {\em Quantum Measurement and Control}.
\newblock CUP, Cambridge, 2010.

\bibitem{Brun02}
Todd~A. Brun.
\newblock A simple model of quantum trajectories.
\newblock {\em {Am.\ J.\ Phys.}}, 70:719, 2002.

\bibitem{Holevo86}
A.~S. Holevo.
\newblock L{\'{e}}vy precesses and continuous quantum measurements.
\newblock In O.~E. Barndorff-Nielsen {\em et al.}, editor, {\em L{\'{e}}vy
  processes: theory and applications}. Birkh{\"{a}}user Boston, Boston, 2001.

\bibitem{Bardou94}
F.~Bardou, J.~P. Bouchaud, O.~Emile, A.~Aspect, and C.~Cohen-Tannoudji.
\newblock Subrecoil laser cooling and {L\'e}vy flights.
\newblock {\em Phys. Rev. Lett.}, 72:203--206, 1994.

\bibitem{Bardou01}
F.~Bardou, J.-P. Bouchaud, A.~Aspect, and C.~Cohen-Tannoudji.
\newblock {\em {L\'e}vy Statistics and Laser Cooling}.
\newblock CUP, Cambridge, 2001.

\bibitem{Breuer07}
H.-P. Breuer and F.~Petruccione.
\newblock {\em The Theory of Open Quantum Systems}.
\newblock Oxford University Press, Oxford, 2007.

\bibitem{Korner}
T.~W. K{\"o}rner.
\newblock {\em Fourier Analysis}.
\newblock CUP, Cambridge, 1988.

\end{thebibliography}

\end{document}